\documentclass{article}
\usepackage{graphicx}
\begin{document}
\title{ The coherent Raman scattering in astrophysics; application to a new model of 
quasar.
\author{J. Moret-Bailly 
\footnote{Laboratoire de physique, Université de Bourgogne, BP 47870, F-21078 
Dijon cedex, France.
Email : Jacques.Moret-Bailly@u-bourgogne.fr
}}}
\maketitle%\textquotedblleft \textquotedblright 

\medskip
Pacs 42.65.Dr Stimulated Raman scattering, 94.10.Gb Absorption and scattering of radiation, 
98.54.Aj Quasars 

\medskip
\textbf{\textit{Abstract}}

 \abstract{

The "Impulsive Stimulated Raman Scattering" (ISRS), is generally performed using ultrashort laser pulses. It shifts  
the frequency of the laser beam without any change of the wave surfaces and any appearance of extra lines. ISRS  
works with the pulses which make the usual incoherent light provided that the interactive medium is a low  
pressure gas having low frequency Raman transitions, that is an hyperfine spectrum. ISRS has no intensity  
threshold, but at low intensity level it becomes a linear effect, so that the relative frequency shift $\Delta\nu/\nu$ 
is nearly constant in the spectrum. This linear effect named "Incoherent Light, Coherent Raman  Scattering" 
(ILCRS) may be confused with a Doppler effect in astrophysics; it explains all optical properties of the quasars, 
including the debatable ones, in particular the width of the BAL lines, the infrared thermal radiation attributed to 
hot dust, and the  spectral discrepancies attributed to a variation of the hyperfine constant.}

\section{Introduction}

	In the spectrum of a quasar, the Ly$_\alpha$ lines and the often associated UV metal lines are generally 
observed with several redshifts; two families of explanations are proposed: either the absorbing gas is in clouds 
next to the quasar, or it is in the intergalactic space. In the first case, the redshift is produced by a Doppler effect 
which requires speeds relative to the kernel of the quasar so high that they seem unreachable  \cite{Goldreich}, in 
the second case it results from the expansion of the universe, except for the \textquotedblleft  broad 
absorption lines (BAL)\textquotedblright. As the absorption lines are generally sharp, the active medium 
must have precisely defined speeds in the first hypothesis, or be confined in relatively thin intergalactic clouds in 
the second hypothesis; an explanation of these confinements is difficult: either they use classical physics, for 
instance magnetic fields, and they are not fully convincing, or they use unobserved matter. The heating of 
intergalactic matter requires the proximity of hypothetical sources. These problems may be solved, but it remains 
debated properties: proximity of a quasar and a galaxy having usually a lower redshift, explanation of a regularity 
in the distribution of the redshifts.

	Already in 1981, hundreds of papers found difficult to explain all redshifts using only the Doppler effect  
\cite{Reboul}. Thus some physicists looked for an alternative to the Doppler effect, but their proposals did not 
work, some were so fantastic that the search for this alternative is often considered as not serious. Strangely the 
coherent Raman scattering was never proposed while Giordmaine et al. \cite{Giordmaine} had described 
experiments in which the frequency of ultrashort laser  pulses was shifted.

	Since, the non-linear light-matter interaction named \textquotedblleft Impulsive Stimulated Raman 
Scattering"  (ISRS) has been extensively studied \cite{Yan,Weiner,Dougherty,Dhar}. ISRS shifts the frequencies 
of the ultrashort laser pulses, introducing  no blur, either of the laser beam, or of the spectral line.

\medskip

 Coherent Raman scattering is surely the source of a part of the observed redshift. In this paper, we propose an 
interpretation of the spectra of the QSOs, using only the usual matter observed by its Lyman and UV spectra and 
an elementary optical effect deduced from regular spectroscopy.

\medskip

	The sentence \textquotedblleft  Impulsive Stimulated Raman Scattering\textquotedblright is somewhat 
ambiguous, it is applied to two  different light-matter interactions :

- Literally ISRS shifts the frequency of a single laser beam, the balance of energy being provided by an excitation  
or de-excitation of the involved molecules; thus its intensity is limited by the saturation of the medium.

- The second one is a combination of two simultaneous preceding effects into a parametric effect; the energy  
provided by a redshift is exactly compensated by a blueshift, so that the molecules remain in their initial states; as  
the intensity is not limited, this parametric effect hides generally the simple one, so that the acronym ISRS 
generally applies  to it. Planck's laws attributing a temperature to the beams, the hot beam is redshifted, the cold 
beam is blueshifted. If the thermal radiation plays the role of cold beams, its blueshift is an amplification, a 
heating.

\medskip

In section 2, the previous {\it ab initio} study of the coherent Raman scattering of incoherent light 
\cite{Moret1,Moret2,Moret3} is avoided, considering that ILCRS is an avatar of the impulsive stimulated Raman 
scattering. G. L. Lamb  \cite{GLamb} gives a general definition of the ultrashort light pulses used to perform ISRS: 
a light pulse is ultrashort if it is \textquotedblleft shorter than all relevant  relaxation times''. The pulses making the 
incoherent light may be \textquotedblleft  ultrashort'' if the two relevant relaxation times in the gases are long 
enough; these relaxation times are the collisional time and the period of the beats of the incident beam with the 
Raman scattered beam, that is the period of the Raman transition.

\medskip

Ordinary incoherent light is made of nanosecond pulses whose intensity is  much lower than the intensity of laser 
pulses. The theory of ISRS shows however that ISRS has no intensity threshold: it may work with ordinary light; 
section 3 shows  that while conventional ISRS is nonlinear, so that the frequency shifts depend on the intensity 
of the laser pulses, at low light level, it becomes linear, so that the relative frequency shift $\Delta\nu/\nu$ is 
nearly constant in the  spectrum; as it is a qualitative change, the new name \textquotedblleft Incoherent Light 
Coherent Raman Scattering" is justified; ILCRS may be confused with a Doppler effect.

Section 4 studies the intensity if ILCRS.

Section 5 shows that ILCRS may hide absorption lines, so that gases such as H$_2^+$, which are easily  
destroyed by the collisions and are ILCRS active cannot be observed by their absorption.

Section 6 sets applications if ILCRS which are important to study the quasars.

Section 7 reminds shortly properties of quasars, including debatable ones and section 8 uses the results of 
section 6 to explain them.

\section{ISRS using nanosecond light pulses.}%2

When a light pulse reaches a mono- or poly-atomic molecule, its eigenstates are mixed with other states and the  
molecule radiates electromagnetic waves; without collisions, quantum or classical computations are the same for  
all identical molecules on a wave front, so that the phases of the emitted waves are the same, the scattering is  
coherent, that is the wave surfaces are preserved. During the light pulse, if the frequencies of the incident and  
scattered lights are different, their relative phase is a linear function of the time. If collisions restart the excitations  
of the molecules, the phases of scattered light become incoherent: it is the ordinary Raman scattering. If the 
incident and scattered light have the same frequency, the collisions preserve the relation of phase between the 
exciting and scattered light, it is the coherent Rayleigh scattering, that is the refraction. The collisions are usually  
taken into account introducing stochastic phase factors in the off-diagonal elements of the density matrix, not in  
the diagonal which corresponds to refraction.

To avoid the incoherence with long pulses, that is to have few collisions during long light pulses, low pressures 
are necessary. Consequently the light-matter interactions are weak, long light paths in the gas are necessary.

\medskip

The sum of two sine waves of close frequencies is a sine wave modulated at the difference of frequencies; 
conversely, the modulation of a sine wave allows to split it into components. If the modulated wave is limited near 
a maximum or a minimum of the envelope, the modulation is negligible so that a single frequency may be detected; 
a trivial computation shows that the sum of the two waves is a single wave whose frequency is intermediate, in 
proportion of the amplitudes \cite{Moret2, Moret3}.

While ILCRS may be powerful, a coherent scattering with a larger Raman shift cannot rise well visible lines: the  
refraction makes the speed of propagation of a discrete Raman line different from the speed of the exciting line;  
thus, the phases of the fields scattered at a distance \textquotedblleft length of coherence'' are opposite where the 
scattered fields  meet; these fields cancel
\footnote{ The beams are supposed wide,  forbidding a notable diffraction which, with laser beams, allows a 
powerful scattering at an angle depending on the ratio of the indices of refraction.}.

To fulfil this condition with 10 nonoseconds light pulses, the Raman period must be larger than 20 nanoseconds,  
that is the Raman transition must correspond to radio-frequencies. Such frequencies are common in heavy  
molecules.  In light molecules, these frequencies are found mostly in  hyperfine structures whose source is either 
an odd number of electrons in a polyatomic molecule, or a Stark or  Zeeman structure in any mono- or polyatomic 
molecule.

\section{Coherent Raman scattering of ordinary incoherent light (ILCRS) as an 
avatar of ISRS.}%3

ISRS is a nonlinear scattering: the scattered amplitude is proportional to the square of the incident amplitude, so 
that the frequency shift is proportional to this amplitude. The computations of ISRS show that, however, it has no 
intensity threshold. As usual in quantum electrodynamics, the amplitude of an electromagnetic field includes the 
zero point field: the square of the amplitude of the electric field in ISRS writes $\widehat E^2 = (E_0 +E)^2$ where 
the zero point field $E_0$ has the same phase than the conventional, usually observed field $E$. For low light 
levels, $E^2$ may be neglected; as the zero point field is nether absorbed or scattered, $E_0^2$ is not taken into 
account; the modulus of $E_0$ being constant in the average, the scattered field which is proportional to $2E_0 
E$, appears proportional to $E$, with the instantaneous phase of $E_0$ and $E$. Thus the frequency shift does 
not depend on the intensity of the beam. The scattered field is stimulated by the zero point field, that is it is 
spontaneous, but coherent in the absence of collisions.

\medskip
In short, the properties of ILCRS are:

- ILCRS works only in very low pressure gases.

- As the light pulses must be shorter than the period corresponding to the virtual Raman transition, ILCRS 
works only if the gas has very low energy transitions, that is, generally, if it has an hyperfine structure.

- While in ISRS the frequency shifts depend on the intensity, they do not in ILCRS. An elementary 
computation shows that the small variation of the relative frequency shift $\Delta\nu/\nu$ in ILCRS results 
from a dispersion of the spectroscopic constants of the gas.

\section{Order of magnitude and dispersion of ILCRS.}%4

ILCRS requires so low pressures, thus so long paths to produce a visible shift that it seems almost impossible to  
measure it in the labs; an other problem is that among ILCRS active, light molecules, only NO and molecules 
perturbed  by a Zeeman effect are stable. An observation of ISRS with nanosecond pulses involves the same 
molecular constants than ILCRS; the difference which results from the induction of the scattered light by the zero 
point field rather than by the a powerful field is easily computed. Thus it seems possible to obtain the ISRS 
parameters of NO with powerful nanosecond laser pulses and a long cell, then deduce the ILCRS parameters of 
this gas.

\medskip
 Theoretical computations are difficult because they involve the parameters of lots of Raman transitions of 
molecules which, even H$_2^+$, are not well known. Up to now, we have done only a very rough evaluation of 
orders of  magnitude, making, to obtain easily understandable results, the unwarranted hypothesis that the 
cosmological  redshift results from ILCRS in an homogenous intergalactic gas.

\medskip
Suppose that the active model molecules have only two low-lying levels and one high such that the two Rayleigh  
and the two Raman probabilities of scattering are the same. A classical computation \cite{Moret1}  leads to a 
needed number of molecules

\begin{equation}
N \approx \frac{16\pi\epsilon_0mk}{he^2}\Bigl(\frac{F}{f}\Bigr)^2H_0T
\end{equation}

where $m$ is the mass of the electron, $e$ its charge, $F$ and $f$ the frequencies of the low and high energy  
transitions, $H_0$ the Hubble constant and T the temperature of the gas. Choosing $T$ = 2,7K and $F/f = 10^6$,  
the number of molecules $N$ is 2 per litre. This result is evidently extremely rough, an error of  several orders of 
magnitude is possible.

\medskip
Suppose that  two spectral lines of frequencies $\nu_1$ and $\nu_2$ are shifted by ILCRS in an homogenous gas  
down to frequencies $\nu'_1$ and $\nu'_2$, so that $\nu'_1<\nu'_2<\nu_1<\nu_2$. The conditions of the redshift  
are exactly the same for both lines between $\nu'_2$ and $\nu_1$, so that the measure of $\nu'_1$ and $\nu'_2$  
can only give an information on the dispersion of the redshifts between the intervals $(\nu'_1,\nu'_2)$ and  
$(\nu_1,\nu_2)$. If the redshifts are much larger than the difference of the eigenfrequencies of the lines, it is 
difficult to observe the dispersion.

\section{Propagation of light in an low absorbing, ILCRS active medium.}%5

The frequencies of a light beam are shifted while the absorption writes lines into the spectrum; an absorption line 
moves in relation to the spectrum brought by the light; thus the width of the lines is equal to the frequency shift; 
if the number of lines is large, as in the spectroscopy of vibration-rotation spectra of polyatomic molecules, the 
lines are mixed, the  absorption is almost uniform, the molecules cannot be detected.

If the pressure of the gas is increased, two behaviours may appear : 

- if the molecules are slowly or not destroyed by the collisions (NO, OH, NH$_2$...) ILCRS disappears before a 
destruction of the molecules, so that the  absorption spectrum appears.

- on the contrary, if the molecules are much sensitive to collisions, as H$_2^+$, they are destroyed before 
ILCRS  disappears. Thus H$_2^+$ cannot be detected in a nearly stable gas.

\medskip
While a line is shifted from frequency $\nu$ to $\nu-\Delta\nu$, it loses the fraction $\Delta\nu/\nu$ of its initial 
energy. This energy amplifies the thermal radiation, including the zero point field; generally the zero point field is 
preponderant, so that the amplified field is isotropic. Thus, if a cloud contains molecules active for the ILCRS, 
while it redshifts the hot beams without blur, it radiates a thermal field in all directions; if the cloud does not 
absorb, its behaviour differs from the behaviour of a cloud of dust, because the number of hot photons is not 
changed; but if the gas absorbs, and if the redshift is large enough to rub the spectral lines, it is very difficult to 
measure a difference.

Neglecting ILCRS, many dusty clouds are observed; near very bright objects, their existence is debatable because 
the grains of dust may be repulsed by the pressure of radiation and corroded by the plasma which surrounds 
them; thus, probably, many apparently dusty clouds are clouds containing only ILCRS active gases.

\section{Propagation of light in a plasma and a magnetic field} %6

Suppose that a light beam propagates in a low pressure plasma of atoms, along an axis $Ox$; in a magnetic field, 
the atoms have an hyperfine Zeeman structure, so that they shift the frequencies by ILCRS. Suppose that the 
static magnetic field $B$ has a zero value at point $O$; near $O$, with a first order approximation, $B$ is a linear 
function of $x$, so that the Zeeman splitting is generally a quadratic function of $x$. Thus the redshifting power 
of the gas is proportional to $x^2$, and to the density of atoms $P$; if $\nu_1$ is the frequency of a spectral 
element for $x=0$, its frequency $\nu$ at $x$ may be written :
\begin{equation}
\nu=\nu_1-\int{3bPx^2{\rm d}x}=\nu_1-bPx^3
\end{equation}
where $b$ depends on the gas.
Supposing that there are nearly no collisions, the gas has a Doppler lineshape, so that the  fraction d$I$ of the 
intensity $I$ of the spectral element absorbed at $x$ may be written:
\begin{displaymath}
{\rm d}I=-kPI\exp(-a(\nu-\nu_0)^2){\rm d}x=
\end{displaymath}
\begin{equation}
=-kPI\exp(-a(\nu_1-bPx^3-\nu_0)^2){\rm d}x \label{raie}
\end{equation}

In the observed spectra, a single line is seen in different places corresponding to the zeros of the magnetic field, 
exactly as if  sheets of absorbing gas moved with various speeds between a white source and the observer. 
Figure 1 shows, for various values of $b$, including $b=0$, the result of a numerical computation of equation 
\ref{raie} for a low absorption, that is considering $I$ as a constant in the second member; the width of the line at 
half intensity is not much increased, but this \textquotedblleft ILCRS line'' has big feet. 

\medskip
%******************
\begin{figure}
\begin{center}
\includegraphics[height=5 cm]{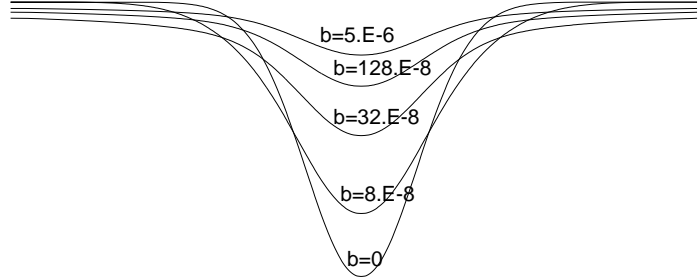}
\end{center}
\caption{ Comparison of a pure Doppler profile (b=0) with the profile resulting from the decrease of ILCRS near a 
zero of the magnetic field, without saturation ( Integration of equation \ref{raie} with I constant in the second 
member).}
\end{figure}
%*******************

Figure 2 shows the result of the computation for large absorptions, such that saturation appears; to allow a 
comparison simple Doppler- and saturation-widened lines are drawn in the lower part of the figure.

%******************
\begin{figure}
\begin{center}
\includegraphics[height=7 cm]{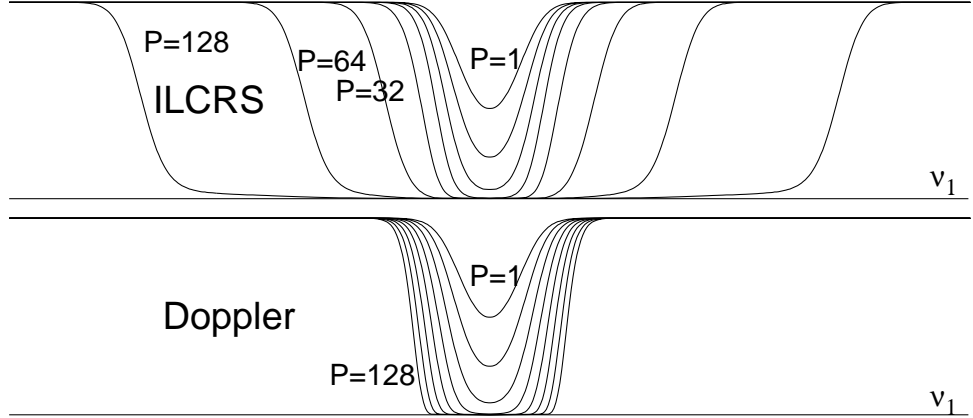}
\end{center}
\caption{ Comparison of the saturation broadening of a pure Doppler profile with the ILCRS profile deduced from 
a numerical integration of equation \ref{raie}. Parameters:  $a$=0.003; $b=8.10^{-8}$; $k$=0,7; $P$ increases by 
steps of a factor 2; $\nu_1-\nu_0$ varies from $-500$ to $+500$.}
\end{figure}
%*******************

\section{Properties of many quasars.}%7
The observations show that, generally:

a) The Lyman absorption spectrum of the quasars is produced by a low pressure gas made of light atoms; the 
temperature is generally higher than 10 000K.

b) Each Lyman line is observed with several redshifts.

c) Most Lyman lines are sharp.

d) Some quasars have broad absorption lines (BAL); their redshifts are lower than the redshift of the sharper 
emission lines, larger than the redshift of the sharp absorption lines.

The following properties are debatable:

e)	The thermal spectra of the QSOs, hotter for BALQSOs \cite{Omont}, is attributed to hot dust.

f)	The density and the temperature of the plasma increase with the redshift \cite{Hui}.

g)	The quasars are next to galaxies whose redshifts are lower \cite{Arp}.

h) The values $z$ of the redshifts verify rules such as $\Delta {\rm ln}(1+z)=0.206$ \cite{Arp2,Narlikar}

i)	The relative frequency shifts are {\it nearly} constant in the spectrum \cite{Webb}.

\section{Simple model of the outer part of a quasar.}%8
Suppose that the quasar is made of a dense, spherical kernel surrounded by a thin atmosphere, origin of the 
emission lines and by a halo heated by the kernel. The next, sister galaxy (Arp's hypothesis) has not such a halo. 
The observation of the polarisation, in the microwave domain, shows the existence of a magnetic field.

In a magnetic (or electric field), the Zeeman (or Stark) effect induces a hyperfine structure in the atoms : the 
plasma around the quasar becomes active for ILCRS. A light beam propagating from the kernel to the earth is 
mostly redshifted during its propagation in the halo. Section 5 shows that the spectrum seems having absorption 
lines only in places which correspond to zeros of the magnetic field. Near the kernel, the emission lines are added 
to the continuous spectrum, they have the maximum redshift. The energy lost by the redshift heats the thermal 
spectrum, simulating heated dust.

The redshifted frequencies do not depend on the speed of a sheet of gas, or the position of an absorbing cloud, 
they detect the zeros of a static field.
In the lower region of the halo of a BALQSOs, the density of atoms is large enough to broaden the ILCRS lines by 
a saturation, as shown in figure 2; elsewhere the lines are sharp. 

Thus, observations a) to e) are explained; f) is a simple consequence of the decrease of temperature and pressure 
with the distance to the QSO; as the sister galaxy has no redshifting halo, the explanation of g) is immediate.

Supposing that the variation of the magnetic field obeys a similar rule for all quasars seems reasonable; it would 
explain h).

\medskip
The difference of the eigen-frequencies of two lines is known from spectroscopy; the difference of the observed 
frequencies appears generally true to the difference of frequencies computed with the hypothesis of a constant 
relative frequency shift $\Delta\nu/\nu$; this usual verification is a strong argument for an interpretation of the 
redshifts by Doppler or expansion effects; as all optical effects, ILCRS is subject to dispersion; section 4 shows 
that the consequence of this dispersion is small.

Using good spectrometers and good calibrations of the spectra, Webb et al. \cite{Webb} found a small variation 
of $\Delta\nu/\nu$ which they considered as a consequence of a variation of the fine structure constant. It is 
simpler to suppose that the discrepancy corresponds to the genuine dispersion ILCRS. 

\section{Conclusion}%9
While optical coherence is taught to all physicists, while we see laser beams almost every day, while ISRS is 
studied for more than thirty years, it is incomprehensible that Raman coherent effects are ignored in astrophysics. 
Low pressure clouds, containing ILCRS active molecules such as OH, NH$_2$, make surely a part of the observed 
redshifts; a precise evaluation of this part requires hypothesis about invisible galactic and intergalactic gases and 
spectroscopic measures and computations, but the rough evaluation forbids to ignore ILCRS {\it a priori}.

With an exception to the origin of radio and microwave frequencies, all properties, sure or debated, of the quasars 
are explained by the hypothesis of a magnetic field in a plasma next to the quasar, plasma whose properties are 
found from the Lyman and associated UV spectrum. The theory is much simpler than the conventional theory, it 
needs only usual matter and regular spectroscopy.

The present hypothesis on the quasars should be refined by a very complicated study of Raman spectroscopy in 
atoms perturbed by Zeeman or Stark effects, and by a  precise study of the magnetic fields near the quasars; but 
the hypothesis gives so remarkably simple an explanation of all optical observations, with only elementary 
hypothesis and regular physical concepts, that it has a large probability of being reliable.

\end{document}